\documentclass[]{tGIS2e}

\usepackage{epstopdf}
\usepackage{enumerate}

\usepackage{enumitem}
\usepackage{hyperref}

\begin{document}

\doi{10.1080/1365881YYxxxxxxxx}
\issn{1362-3087} \issnp{1365-8816} \jvol{00} \jnum{00} \jyear{2015} 

\markboth{William Lee Croft, Wei Shi, J\"{o}rg-R\"{u}diger Sack and Jean-Pierre Corriveau}{International Journal of Geographical Information Science}


\title{Geographic Partitioning Techniques for the Anonymization of Health Care Data}

\author{William Lee Croft$^{a}$$^{\ast}$\thanks{$^\ast$William Lee Croft (Corresponding author). Email: lee.croft@carleton.ca
\vspace{6pt}}, 
Wei Shi$^{b}$$^{\ast}$\thanks{$^\ast$Wei Shi (Corresponding author). Email: wei.shi@uoit.ca
\vspace{6pt}},
J\"{o}rg-R\"{u}diger Sack$^{a}$$^{\ast}$\thanks{$^\ast$J\"{o}rg-R\"{u}diger Sack. Email: sack@scs.carleton.ca
\vspace{6pt}} 
and Jean-Pierre Corriveau$^{a}$$^{\ast}$\thanks{$^\ast$Jean-Pierre Corriveau. Email: jeanpier@scs.carleton.ca
\vspace{6pt}}\\\vspace{6pt}  
$^{a}${\em{School of Computer Science, Carleton University, Ottawa, Canada}} \\
$^{b}${\em{Institute of Technology, University of Ontario, Oshawa, Canada}}
\\\vspace{6pt}\received{v1.0 released May 2015} }
\maketitle

\begin{abstract}
Hospitals and health care organizations collect large amounts of detailed health care data that is in high demand by researchers. Thus, the possessors of such data are in need of methods that allow for this data to be released without compromising the confidentiality of the individuals to whom it pertains.  As the geographic aspect of this data  is becoming increasingly relevant for research being conducted, it is important for an \emph{anonymization} process to pay due attention to the geographic attributes of such data. In this paper, a novel system for health care data anonymization is presented. At the core of the system is the aggregation of an initial regionalization guided by the use of a Voronoi diagram. We conduct a comparison with another geographic-based system of anonymization, GeoLeader. We show that our system is capable of producing results of a comparable quality with a much faster running time.\bigskip

\begin{keywords}Data Anonymization; Geographic Partitioning; Health Care
\end{keywords}\bigskip

\newpage
\end{abstract}

\section{Introduction}
For researchers in the health care domain, detailed data sets are essential. As such, there is a high demand for the availability of this type of data. However the sensitive nature of the data prevents hospitals and organizations in possession of such data from releasing it in its original state. In particular, government policies impose restrictions on what can be released in order to protect the confidentiality of the patients and respondents to whom the data pertains \cite{2,4,3,5,7,1,6}.

In order to render a data set safe for release, any directly identifying information such as the names of patients must be removed. However, there are methods that a malicious party can make use of in order to re-identify the remaining data \cite{3}. Further modification of the remaining attributes is therefore necessary in order to sufficiently protect the data for release. There are many existing systems that address this \emph{anonymization} problem; however few systems place an emphasis on the geographic information contained within these data sets. During the process of rendering a data set safe for release, information will invariably be lost. The importance of the data that is lost varies with respect to the context of the intended research drawing on the resultant anonymised data. For some studies, such as those that observe the propagation of diseases over geographic areas, the geographic information in the data set is of high importance \cite{68}. Research fields such as spatial epidemiology require high precision geographic information in order for researchers to work effectively
\cite{9,8,10}. However, the release of this high precision geographic information creates a much higher risk for the disclosure of confidential information due to the fact that individuals in small geographic areas are easily re-identifiable. It is therefore necessary to devise methods that can allow for greater precision in geographic information to be disclosed while still protecting the confidentiality of the individuals in the released data.

Existing geographic-based anonymization methods have a tendency to cause a greater loss of geographic information than necessary. For example,  cropping \cite{41,48}  can cause a high loss in geographic precision, and the suppression of at-risk regions \cite{34,27}  can produce a heavily censored data set. Any loss in the geographic information has a negative impact on the quality of the research that is conducted using the resultant data \cite{47}.

In this paper, we present a geographic-based system for health care data anonymization, which employs a process of geographic partitioning guided by a Voronoi diagram \cite{99}. Our system, Voronoi-Based Aggregation System (VBAS), achieves k-anonymity \cite{13,11} on a data set through the generalization of geographic attributes used in combination with the suppression of outlying records. VBAS is able to reduce the loss of geographic information in order to produce data that is more useful for researchers. By aggregating small regions of fine granularity into larger regions that satisfy a specified anonymity requirement, we are able to produce regions that maintain a high level of geographic precision.

We compare VBAS to GeoLeader\footnote{A very quick introduction to this anonymization system can be found at http://www.slideshare.net/LukArbuckle1/geo-leader-tophc-poster-2013-v2. It should be noted that there are several other systems called GeoLeader that do not pertain to data anonymization.} \cite{63}. More specifically, through tests run on synthetic data sets generated from Statistics Canada data, we compare actual implementations of VBAS and  GeoLeader in terms of their speed, their ability to reduce information loss, and the compactness of regions that they create. We demonstrate that VBAS is capable of producing results of similar quality with a few distinct advantages. VBAS runs significantly faster, provides a guarantee of k-anonymity which GeoLeader does not, and is produces aggregated regions which are slightly more compact.

The choice of GeoLeader proceeds from two considerations: First, we had access to a running implementation of it, and second, the output of this implementation could be used to compute metrics that are also relevant to VBAS. In contrast, an implemented geographic-based solution depending on masking \cite{111} or linear programming \cite{48} was not readily available and, more importantly, would not allow us to directly reuse the metrics we consider (namely reduction in geographic precision).

\section{Literature Review}
\subsection{Anonymizing Health Care Data}
When a health care data set is to be released, it is necessary to guard against the risk of re-identification. In other words, the released data must first be de-identified in such a way that it should not be possible to use it to associate any confidential information with a specific individual \cite{4,3,5,7,1,6}. The general process used to de-identify a data set is to remove all directly identifying attributes such as names and identification numbers and then modify the demographic-type attributes, referred to as quasi-identifiers, so that they cannot be used to re-associate any directly identifying attributes with the records of the data set \cite{4,5,7,6}.

\begin{figure}
	\caption{Equivalence Classes}
	\label{fig:equiv}
	\centering
	\includegraphics[width=0.7\columnwidth]{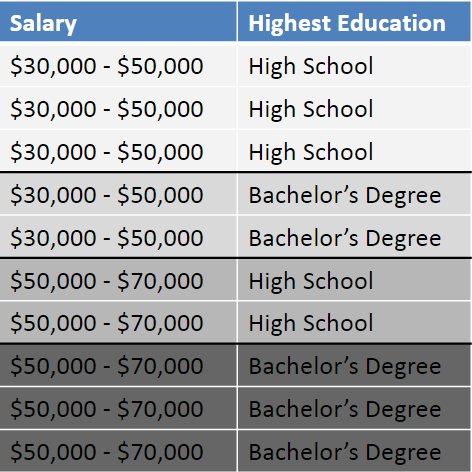}
	
	Each group of records highlighted in a different shade is an equivalence class.
\end{figure}

The main goal of de-identification is therefore to reduce the risk of re-identification to an acceptably low level. There are, however, no standardized methods for measuring the actual risk of re-identification. As such, it is necessary to find factors that can be used to analyze the risk. One such factor that is commonly used is the distinctness of the records in a data set. Records are compared to each other based on their quasi-identifier attribute values. The combination of these values in a record determines an equivalence class into which the records falls. Thus, any record that does not have any other records in a data set that completely match its quasi-identifier attribute values is considered to be unique, making it the sole member of its equivalence class. An example of equivalence classes can be seen in the sample data set of Figure \ref{fig:equiv}. Records which are unique or which belong to an equivalence class with few members are considered to have a higher risk of re-identification as a party can cross-reference the quasi-identifiers with other publicly available data sets in order to associate directly identifying information with the records. The higher the cardinality of an equivalence class is, the harder it becomes for useful information to be discerned via cross-referencing as it becomes necessary for the party attempting the cross-reference to make guesses with lower probabilities of being correct due to the records being indistinguishable from each other \cite{4,3,5,6}.	

Since the reduction in the distinctness of records can be used as means of protection for confidentiality in these data sets, methods have been devised to lower the distinctness of records. The two techniques that are most commonly used to do this are generalization and suppression. Generalization is typically applied globally to modify the granularity of the response categories of quasi-identifiers \cite{14,12,15,13,11}. By coarsening this granularity, the number of equivalence classes is reduced which in turn causes higher cardinalities in the remaining equivalence classes. Suppression is generally applied at a local level to remove individual records from the data set \cite{14,12,15,13,11}. This is typically done in cases where outlying records exist in equivalence classes of low cardinalities. In such cases, further application of generalization, which affects all records, may be more heavy-handed than necessary, thus, suppression is a more reasonable choice.

These methods used to reduce the distinctness of records are generally applied in a structured fashion in order to produce a data set that conforms to some type of guarantee about its level of protection. A commonly used guarantee is k-anonymity \cite{14,12,15,13,11}. In order for a data set to be considered k-anonymous, every equivalence class must have a cardinality greater than or equal to k, which is a user-specified value. By ensuring that all equivalence classes have a sufficiently high cardinality, the risk of disclosure via cross-referencing can be considered sufficiently low.

\subsection{Geographic-Based Approaches to Anonymity}
One possible approach to de-identification is to focus on the geographic aspect of the data. Since the data can be protected by ensuring that levels of distinctness are sufficiently low, a large enough region can be used to provide a sufficiently large group of records, thus decreasing the levels of distinctness in the records. If, however, the records are grouped into small regions, the inclusion of the geographic attribute value will hinder the creation of high cardinality equivalence classes. The level of granularity used for geographic precision is therefore an important factor to consider. The reduction of geographic precision is essentially a form of geographic generalization which can be used to control the sizes of equivalence classes. The other quasi-identifiers still play a part in determining the equivalence classes as well, thus, if they are too finely grained to start with, this will not be effective.

Studies done on the relationship between the geographic aspect of data sets and the levels of distinctness have shown that when a population is sufficiently large, the data will have an acceptable level of anonymity if the entire population uses the same geographic attribute value \cite{29,28,27}. In other words, the reduction of geographic precision can be used as means to achieve anonymity. Despite this, it is still desirable to maintain as much geographic precision as possible for research purposes, thus, there is a trade-off.

A simple approach to employ this concept is to determine a population cutoff size for a data set. This cutoff size indicates what population size must be exceeded in order for a region to have an acceptable level of anonymity. Any region that does not exceed this population size will then have all of its records suppressed. Examples of this approach can be seen with data sets produced from census surveys such as in the United States where The Bureau of Census employs a 100,000 population size cutoff \cite{27}. Similarly, Statistics Canada uses a 70,000 population size cutoff for their Canadian Community Health Survey \cite{30} and the British Census uses a 120,000 population size cutoff \cite{31}.

The use of cutoff sizes to suppress small regions, however, suffers from some drawbacks. The suppression of these small regions has the potential to cause a very large number of records to be suppressed and can result in a highly censored data set. Additionally, each data set must be studied individually in order to manually determine an appropriate cutoff size. It is not possible to select a single standardized cutoff size since the quasi-identifiers of each data set influence what size will be appropriate for the data set. The issue of the cutoff size can be addressed by a method that dynamically computes a cutoff size for a given data set based on its quasi-identifiers \cite{34,35}. The problem issue of over-suppression, however, remains.

An alternative approach is to reduce geographic precision by widening the areas referred to by the geographic attributes of the data set. Cropping \cite{41}, for example, removes the last three characters from postal codes in order to refer to much larger geographic areas. A geographic generalization hierarchy can also be applied to reduce precision. An example of such a hierarchy would be to generalize from the level of postal codes to cities, and from cities to provinces. These methods, however, have the potential to cause a much greater loss in geographic precision than necessary. In order to better preserve the precision, a more finely grained method of widening areas is required.

A greater control in the reduction of precision can be achieved through the aggregation of regions rather than by following a hierarchy to reduce precision. This allows for regions to be grown in much smaller increments. In order to do so effectively, it is necessary to know at what point to stop the growth of the regions and to determine which regions are best suited to be aggregated together. A recent system, GeoLeader \cite{63}, takes this approach by first computing a dynamic cutoff size for the input data set and then running an iterative process of aggregation. At each iteration of this process, all adjacent regions are evaluated using a set of criteria to determine which candidates are the best choice for aggregation. The process stops when all regions have a population above the computed cutoff size. It should be noted that GeoLeader offers only the guarantee that the aggregated regions will have a population above the cutoff size; it does not enforce k-anonymity on the resultant data set.

\subsection{Data Utility Metrics}
During the anonymization of a data set, information is lost when the quasi-identifiers undergo modification. This loss in information reduces the utility of the data for researchers. It is therefore useful to be able to measure how much information has actually been lost. Although there are no standardized measures for this, various metrics have been proposed.

\subsubsection{Discernibility Metric}
A discernibility metric was introduced in \cite{67} and has been applied in other systems \cite{26,12} as well. This metric assigns a penalty to each record of the anonymized data set proportional to the number of other records from which it is indistinguishable. Despite the fact that this indistinguishability is useful for the protection of privacy, it also reduces the utility of the data.

\subsubsection{Non-Uniform Entropy Metric}
A non-uniform entropy metric has also been applied to measure information loss. This metric, first applied in \cite{25} and also used in \cite{12}, calculates the loss in information based on the probability of correctly guessing original attribute values of records given their anonymized values. It works under the assumption that a greater amount of information is lost in cases of uniform distributions of attribute values than in cases of non-uniform distributions as it is harder to guess the original values in these cases. The total value representing the amount of lost information is calculated based on the application of this probability across each attribute of each record in the data set. The higher the calculated value, the greater the degree of information loss.

\section{VBAS Details}
The general idea behind VBAS is to achieve anonymity on a data set through aggregation run on an initial regionalization of fine granularity. The selection of regions to aggregate is guided by the use of a Voronoi diagram \cite{99}. Voronoi diagrams are employed in many different fields and applications \cite{98} as tools or building blocks in the design of algorithms. The Voronoi diagram, which takes a set of sites (point locations) as input, provides a means to efficiently divide the plane into convex regions where each region corresponds to one of the input sites. Each Voronoi region consists of the area in the plane that is closer the region's site than to any other site. This representation uses linear space in the number of sites. Since the latter will be much smaller than the number of initial regions, this uses a relatively small amount of storage space.

The system takes two files as input: one that contains information about the initial regionalization and one that contains information about the data set to be anonymized. The initial regions must be represented in the plane as a point set, either by using coordinates provided in the input file or by computing the centroids of the regions. The Voronoi diagram can then be used to determine the regions to aggregate together by creating groupings of the initial regions based on the Voronoi regions in which their point representations fall. These groupings of regions indicate the initial regions which will be aggregated together.

In order to produce an aggregation with desirable qualities, it is important to carefully select the number of sites to supply for the Voronoi diagram as well as the locations at which to place the sites. The system is thus broken up into four main components:
\begin{itemize}[noitemsep,nolistsep]
	\item Approximating an appropriate number of aggregated regions
	\item Selecting locations at which to place Voronoi sites
	\item Constructing the Voronoi diagram and performing aggregation
	\item Rating the aggregation
\end{itemize}
\bigskip

\begin{figure*}[!htbp]
	\centering
	\caption{VBAS Screenshot}
	\label{fig:screen}
	\centering
	\includegraphics[width=\textwidth]{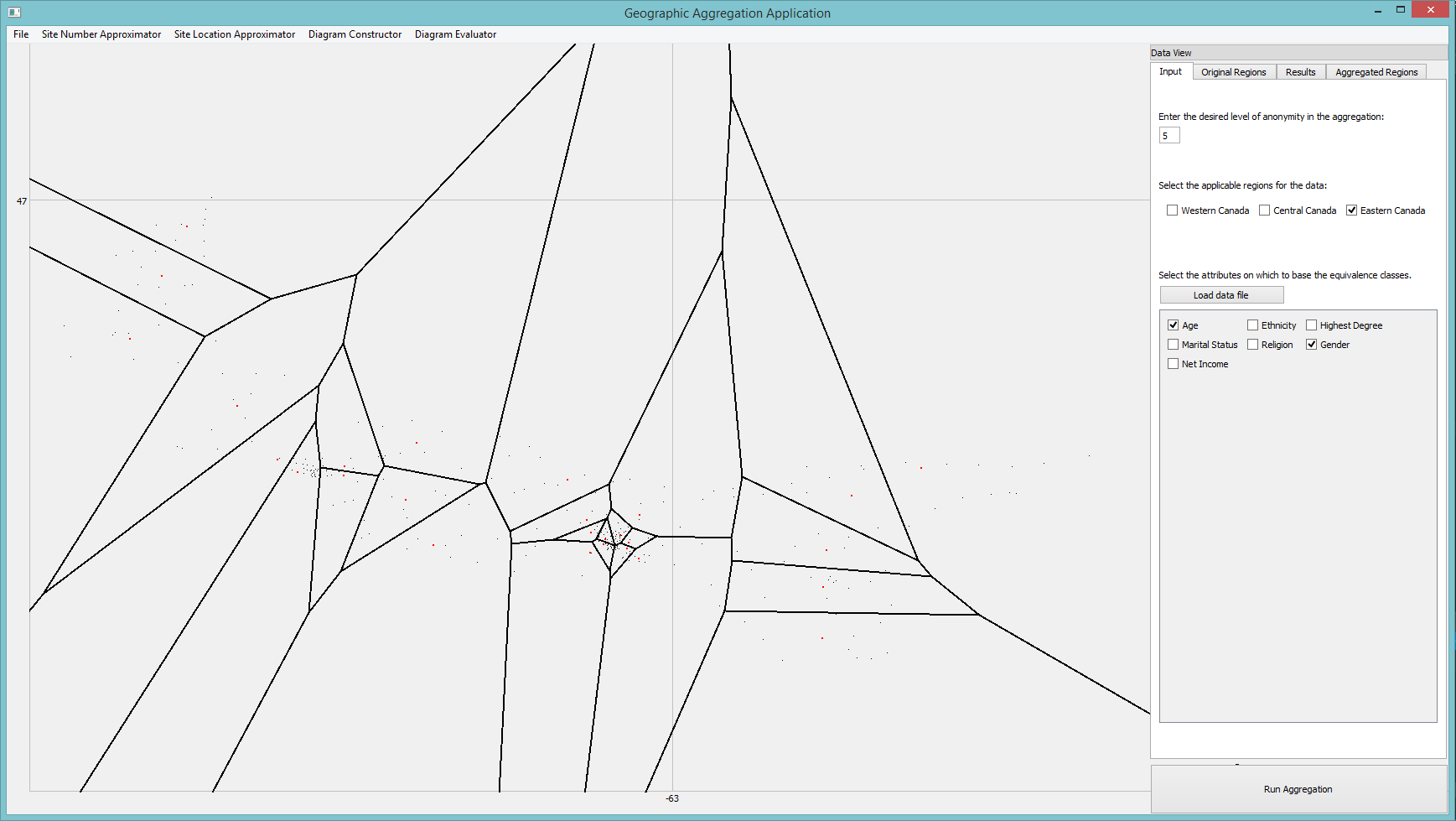}
\end{figure*}

A screenshot of the application with an aggregation on display can be seen in Figure \ref{fig:screen}. The system components are designed so that they can be supplied with different approaches to accomplish the work that they must do. This is done to allow for ease of configurability and to provide the ability to easily test different combinations of approaches. For the comparison conducted in this paper, we have selected a single set of appropriate approaches.

\subsection{Site Number Approximation}
The selection of the number of sites has a large influence on the levels of anonymity in the aggregated regions and must therefore be carefully approximated. Each site will produce a single Voronoi region and by extension a single aggregated region. Therefore, if the approximation is too low then too few aggregated regions are created resulting in a great loss of geographic precision. If, however, the approximation is too high, there will be many aggregated regions that may not attain sufficient levels of anonymity.

Our tests have shown that while there is a need for the approximation to be balanced, there is some room to allow for the selection of the number of sites to be adjusted based on user requirements. For example, if a user prioritizes geographic precision over the reduction of suppressed records, a higher approximation can be made to accommodate this.

Despite the ability to tune the results in this way, the approximation of the number of sites is excluded from the comparison conducted in this paper. In order to provide a more accurate comparison between VBAS and GeoLeader, we match the number of aggregated regions created by the two systems. This allows for us to more closely compare their abilities to reduce the loss of geographic information during aggregation. Although we can compare the two systems when they each make their own decision for the number of aggregated regions to create, this is of little interest. As observed in our previous tests, the adjustment of this approximation reveals a tradeoff between the reduction of information loss and the reduction of suppression. That is, either one may be sacrificed in favor of the other. With the ability to adjust these approximations in our system through the selection of different site location approaches, there is little to be discerned from such a comparison.

\subsection{Site Location Selection}
The selection of the locations for the sites determines the shape and size of the aggregated regions that will be created as well as the levels of anonymity in those regions. These locations must therefore be carefully selected in order to create regions with appropriate levels of anonymity while attempting to reduce the loss of geographic information. A fundamental characteristic of VBAS is it is highly configurable. More precisely, we have experimented with many different strategies for setting its different approximations (as discussed at length elsewhere \cite{112}). In this paper, for site location selection, we employ what we call a \emph{balanced density} approach (B. Density in figures below), which was found to produce very good results for both reduction of suppression and reduction of information loss, and was able to do so very quickly (\emph{Ibid.}). 

This approach divides the plane into a number of cells equal to the number of sites to place. The goal is to create these cells such that they will each have roughly the same population across the initial regions that fall within the cells. A single site is then allocated to each cell. The reason for doing so is to try to match the distribution of the sites with the distribution of the population in order to create aggregated regions that will have roughly the same population levels. Throughout this approach, the boundaries of the cells are never actually drawn. The cells are containers for the initial region points; the concept of boundaries is simply used as an aid for the visualization of the approach. For ease of organization, the cells are grouped together into rows such that all cells in a row have the same upper and lower boundaries (those of the row) and occupy the entire space covered by the row.

The first step is to make an approximation for the number of rows that will be needed. This approximation does not need to be very precise as it will actually be adjusted later; it is simply used as a starting point. We therefore take the square root, rounded to the nearest integer, of the number of sites as the initial approximation. This assumes an even distribution of the population, meaning that the number of cells per row and the number of rows would be roughly the same. The calculation is shown in Equation 1.
\medskip
\\
Let:
\begin{itemize}[noitemsep,nolistsep]
	\item[$r$] be the number of rows
	\item[$s$] be the required number of sites
	\item[$R(x)$] be a function that rounds x to the nearest integer value
\end{itemize}

\begin{equation}
r = R(\sqrt{s})
\end{equation}

With the initial number of rows determined, the ideal population per row is calculated as the total population of the data set divided by the number of rows, rounding this quotient to the nearest integer value. This is shown in Equation 2.
\medskip
\\
Let:
\begin{itemize}[noitemsep,nolistsep]
	\item[$p$] be the total population
	\item[$p'$] be the ideal population per row
\end{itemize}

\begin{equation}
p' = R(\frac{p}{r})
\end{equation}

The next step is to then determine the divisions between the rows. This is done by allotting a population to each row that is as close as possible to the ideal row population. In order to do so, all initial region points are first sorted by their y-coordinates. We then walk across these points starting from the one with the lowest y-coordinate, stopping when the sum of the population across the points that have been passed is greater than or equal to the ideal row population. Since each of the points represents an initial region, they will each have a population associated with them. At this point, it must be decided whether the final point that was passed is best suited for the current row or for the next row. This is determined by comparing the population of the current row with and without the final point. If the population of the row including the point is closer to the ideal population then it is kept. If the population without the point is closer, it is left for the next row. This is shown in Equation 3.
\medskip
\\
Let:
\begin{itemize}[noitemsep,nolistsep]
	\item[$r_{p'}$] be the population of a row just before it passes the ideal population
	\item[$r_{p''}$] be the population of a row just after it passes the ideal population
	\item[$I(r)$] be an indicator function for a row r where its value is 1 when the final point should be included in the row and 0 when it should not
\end{itemize}

\begin{equation}
I(r) = \left\{	\begin{array}{ll}
1  & \mbox{if } r_{p''} - p' \le p' - r_{p'} \\
0 & \mbox{if } r_{p''} - p' > p' - r_{p'}
\end{array}	\right.
\end{equation}

Once the points for the row have been determined, they are stored in a container and the walk continues in order to determine the points of the next row. Since the population of a row increases by intervals corresponding to the population of each point, it is not possible to guarantee that the population of a row will match the ideal population, in fact, it is highly unlikely that it will. This means that due to rows that take on a greater population than desired or rows which take on a lower population than desired, it is possible to run into two different scenarios that require the number of rows to be adjusted.

The first scenario occurs when rows have taken on a greater population than desired resulting in an insufficient remaining population to fill up the rest of the rows. When this occurs the first row that cannot be sufficiently filled will simply take on the entire remaining population. The total number of rows is then adjusted to the number of rows that have been created so far. Any other rows that were to be created originally will simply no longer be used.

In the second scenario, the rows have taken on a smaller population than desired, leaving the final row with a much greater population than intended. In this case, additional rows will be created by continuing the process of walking across the points and creating a new row each time the ideal population is passed. The total number of rows is thus increased to however many rows were created.

In either scenario, the adjustment of the number of rows does not pose any problems in terms of satisfying the requirements of the approach. It is only necessary to create a number of cells equal to the number of sites to be placed and to ensure that the cells have roughly the same population. Each row will have a number of cells assigned to it based on the population of the row. As such, the actual number of rows can be freely adjusted.

Once the rows have been created, each of them must be addressed individually as the number of cells allotted to each row is a function of its population. In order to calculate this number of cells, the population of the row is divided by the total population of the data set to determine the percentage of the total population within the row. The number of sites to place is then multiplied by this percentage and rounded to the nearest integer to determine how many sites should be placed in the row. This number of sites is used as the number of cells to create. The calculations are shown in Equations 4 and 5.
\medskip
\\
Let:
\begin{itemize}[noitemsep,nolistsep]
	\item[$r_p$] be the population of a row
	\item[$r_\alpha$] be the decimal percentage of the total population in a row
	\item[$r_c$] be the number of cells assigned to a row
\end{itemize}

\begin{equation}
r_\alpha = \frac{r_p}{p}
\end{equation}
\begin{equation}
r_c = R(s(r_c))
\end{equation}

The divisions between the cells of a row are made in the same way as the divisions between the rows. The points of the rows are first sorted by their x-coordinates and are then walked across from left to right, creating a division between cells each time the ideal cell population has been passed. Just as with the rows, each cell has its points stored in a container.

Once again, two scenarios must be addressed in which cell populations differing from the ideal populations have caused an inappropriate population to be left for the remaining cells. Now, however, the number of cells cannot be adjusted since the total number of cells must match the number of sites to be placed. To accommodate for this, if a greater population is left for the final cell than the ideal population, that cell simply takes on the entirety of the remaining population. While this has a negative impact in terms of keeping the population of that cell similar to the population of the other cells, it is necessary in order to keep the number cells the same. In the case where the remaining population is too small to fill the cells that must still be created, the entire remaining population is allotted to a single cell. Then the largest cells of the row are split in two until the required number of cells for the row has been reached.

With all of the cells created, one site can then be allocated per cell to be placed at the median of the points in the cell. Equations 6 and 7 show the computation of the median for a cell.
\medskip
\\
Let:
\begin{itemize}[noitemsep,nolistsep]
	\item[$P$] be the set of points in a cell
	\item[$p_i$.x] be the x-coordinate of a point $p_i$
	\item[$p_i$.y] be the y-coordinate of a point $p_i$
	\item[$m.x$] be the x-coordinate of the median
	\item[$m.$] be the y-coordinate of the median
\end{itemize}

\begin{equation}
\large
m.x = \frac{\sum_{p_i \in P}p_i.x}{|P|}
\end{equation}	
\begin{equation}
\large
m.y = \frac{\sum_{p_i \in P}p_i.y}{|P|}
\end{equation}

\subsection{Construction of Geographic Aggregation}
With the site locations selected, these locations can be provided as input to construct the Voronoi diagram \cite{99}. Once the diagram is constructed, the initial regions are grouped together based on the Voronoi region in which their point representation falls. Since the Voronoi diagram is a planar subdivision, point location can be efficiently conducted. The groupings of initial regions determine which regions will be aggregated together into a single new region.

During aggregation, the equivalence classes of the aggregated regions must be determined by combining the members from the equivalence classes of the initial regions being merged together. This is done in order to verify the cardinalities of the resultant equivalence classes. If any equivalence class in an aggregated region has a cardinality less than the user selected value of k for k-anonymity, all records of the equivalence class must be suppressed in order to ensure a sufficient level of anonymity.

\subsection{Evaluation of Aggregation}
The final component of the system is used to evaluate the quality of the aggregation that has been produced. The approach that we apply for this is a grouping of the following measurements:
\begin{itemize}[noitemsep,nolistsep]
	\item Suppression
	\item Compactness
	\item Discernibility
	\item Non-Uniform Entropy
	\item Running Time
\end{itemize}
\bigskip

\subsubsection{Suppression}
Suppression is measured as the total number of records that were suppressed during aggregation. Higher levels of suppression indicate a greater number of equivalence classes of low cardinalities meaning that the aggregation was not effective in achieving anonymity.

\subsubsection{Compactness}
The compactness of the aggregated regions can be used as an indication of how much geographic precision exists in the release data set. More compact regions provide a higher level of precision. This measurement is taken as the sum of the distances between each initial region point and the median of the initial region points in its aggregated region as shown in Equation 8.
\medskip
\\
Let:
\begin{itemize}[noitemsep,nolistsep]
	\item[$R$] be the set of aggregated regions
	\item[$r_{ip}$] be the centroid of the initial regions in the aggregated region r$_i$
	\item[$R_i$] be the set of initial regions in the aggregated region r$_i$
	\item[$r'_{ip}$] be the point representation of the initial region r'$_i$
\end{itemize}

\begin{equation}
\sum_{R'_i \in R}\sum_{r'_i \in R'_i}\left\|r_{ip} \mbox{ } r'_{ip}\right\|
\end{equation}
\bigskip

\subsubsection{Discernibility}
The discernibility \cite{26,12} information loss metric is employed in order to check for overburdened equivalence classes that indicate a greater loss of information than necessary. The calculation for this metric is shown in Equation 9. Higher values indicate a greater loss of information during aggregation. 
\medskip
\\
Let:
\begin{itemize}[noitemsep,nolistsep]
	\item[$E$] be the set of equivalence classes
	\item[$E_i$] be an equivalence class from the set E
	\item[$k$] be the desired level of anonymity
\end{itemize}

\begin{equation}
\sum_{\left( |E_i| \ge k \right) \in E}|E_i|^2
\end{equation}
\bigskip

\subsubsection{Non-Uniform Entropy}
The non-uniform entropy \cite{12, 15} information loss metric is also employed. In order to make this measurement, it is necessary to calculate the probability of correctly guessing the original geographic attribute value of a record given its aggregated value. This is shown in Equation 10.
\medskip
\\
Let:
\begin{itemize}[noitemsep,nolistsep]
	\item[$a_r$] be the original value of the attribute
	\item[$b_r$] be the generalized value of the attribute
	\item[$n$] be the number of entries in the data set
	\item[$I\left( \right)$] be the indicator function
	\item[$R_i$] be original attribute value of the i$^{th}$ entry
	\item[$R'_i$] be the generalized attribute value of the i$^{th}$ entry
\end{itemize}

\begin{equation}
Pr\left( a_r | b_r \right) = \frac{\sum_{i=1}^n I\left( R_i = a_r \right)}{\sum_{i=1}^n I\left( R'_i = b_r \right)}
\end{equation}
\bigskip

The calculation of this probability is applied to each record of the data set in order to produce a measure of the information loss using the calculation shown in Equation 11. Once again, a higher value indicates a higher level of information loss.
\medskip
\\
Let:
\begin{itemize}[noitemsep,nolistsep]
	\item[$R_i$] be original geographic identifier of the i$^{th}$ entry
	\item[$R'_i$] be the generalized geographic identifier of the i$^{th}$ entry
	\item[$n$] be the number of entries in the data set
\end{itemize}

\begin{equation}
-\sum_{i=1}^n \log_2 Pr\left( R_i | R'_i \right)
\end{equation}
\bigskip

\subsubsection{Running Time}
The measurement of running time is simply a measure of how long each system takes from the start of its process to the end.

\section{Comparisons}
To compare VBAS and GeoLeader, we have generated test data through the use of publicly available data sets from Statistics Canada. Using the generated testing set, we have run different scenarios to compare the effectiveness of the systems in the various measurements. All tests were run on a machine using 16 GB of RAM and a 4.01 GHz processor.

\subsection{Generation of Testing Data}
In order to generate the testing data set, the public use microdata file from the 2011 National Household Survey \cite{49} (NHS) and the Canadian dissemination areas data set \cite{50}, both available from Statistics Canada, were used. As required by Statistics Canada's data use regulations, it is stated that the results or views expressed here are not those of Statistics Canada.

We have used the NHS data set to make approximations for the distributions of attribute values across the response categories of a selection of the demographic attributes. Since the data set contains respondent level information for a 2.7\% sample of the Canadian population, these approximations were made in each of the provinces and territories. It was necessary to make the approximation at such a rough level of geographic precision as this was limited by the geographic information available in the NHS data set.

Next, going by Statistics Canada's documentation that indicates that dissemination areas have a population targeted between 400 and 700 \cite{50}, we randomly generated a number within this range for each dissemination area to act as its population and then created an appropriate number of records to fill the dissemination area. Each generated record was assigned values on each selected quasi-identifier attribute by assigning a value from the response categories with a probability corresponding to our approximations of the distribution for the attribute within the province or territory in which the dissemination area exists. We also assigned a geographic attribute to each record indicating its dissemination area in order to provide a finely grained geographic precision. By generating records for each dissemination area in this way, we have produced a testing set for all of Canada.

Since the GeoLeader system requires information about the radii and adjacencies of its initial regions, our tests were limited to Prince-Edward Island (PEI), which is the only province for which we had access to this particular information. As such, we have created a subset of the Canada testing set that contains only the records in PEI.

\subsection{Test Scenarios}
The tests run on the two systems consist of scenarios defined by different selections of quasi-identifiers on which to achieve anonymity. In total, eighteen different scenarios were tested. In each of them, both systems were run using the same PEI data and the results were recorded in terms of the measurements discussed in the evaluation component of our system. In order to record these measurements for the GeoLeader system, we have written a supplemental application that accepts the output of the GeoLeader system and processes it to produce the measurements needed for the comparison.

\section{Discussion of Results}
In all test scenarios, the measurements in suppression, discernibility, and non-uniform entropy were very close in both systems. For the marginal differences that were observed, neither system consistently remained on the same side of the margin. From this, we conclude that the two systems are roughly evenly matched in terms of their abilities in the reduction of suppression and the reduction of geographic information loss.

\begin{figure*}
	\centering
	\caption{Compactness Comparison}
	\label{fig:compactness}
	\centering
	\includegraphics[width=\textwidth]{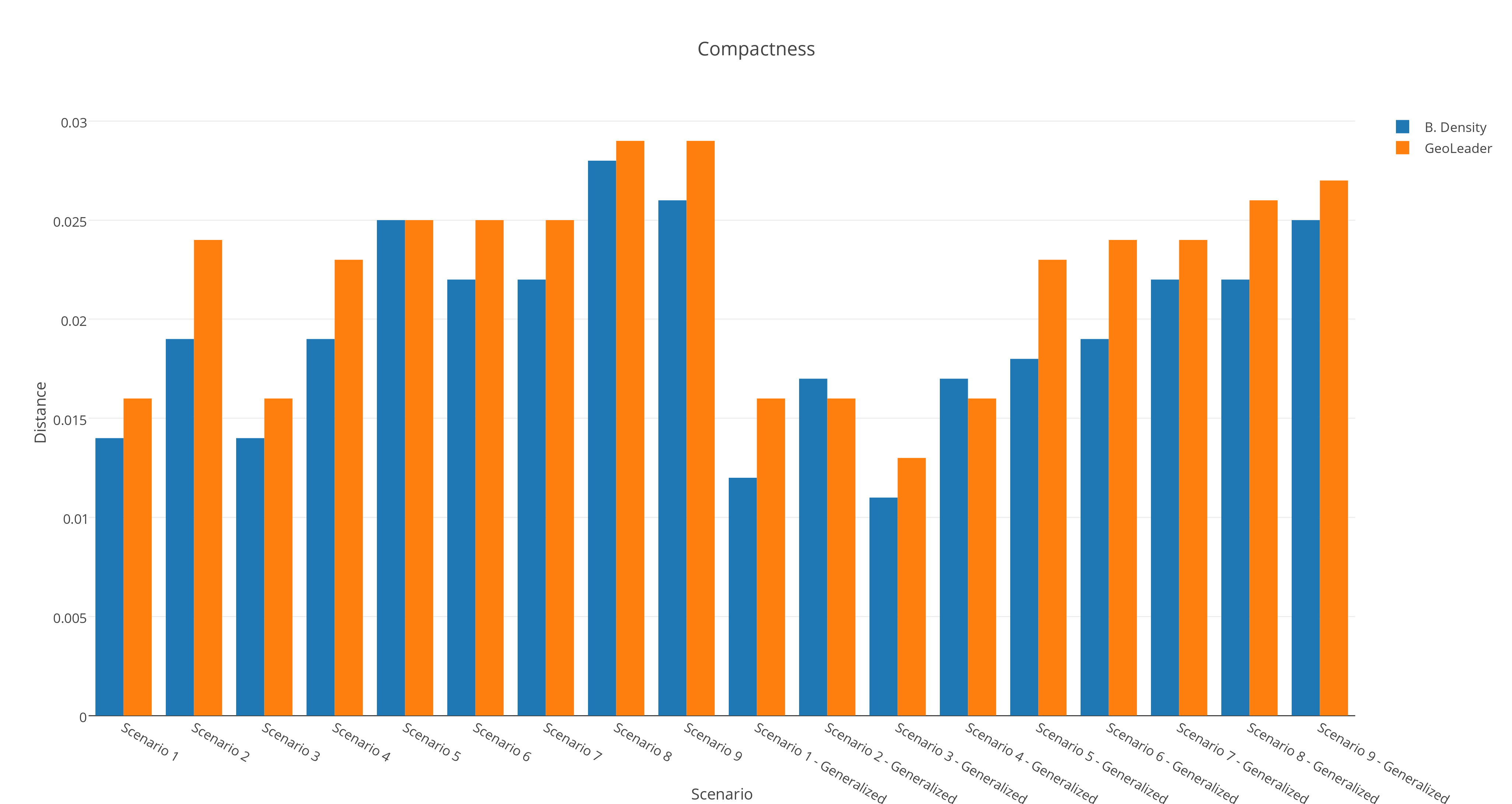}
\end{figure*}

When comparing the measure of compactness, our system consistently outperformed GeoLeader by a small margin. On average, VBAS had a compactness measurement at 85.4\% of that of GeoLeader. A visual comparison of the result can be found in the graph in Figure \ref{fig:compactness}. GeoLeader applies criteria during the selection of regions to merge in order to maintain compact aggregated regions. This ensures that it will not create overly elongated or sprawling regions however it is still susceptible to a loss in compactness due to concavities. Our system has an advantage in this regard resulting from the use of the Voronoi diagram. Due to the property of the Voronoi diagram that guarantees that all Voronoi regions will be convex \cite{99}, our resultant aggregated regions have a high level of compactness.

\begin{figure*}
	\caption{Running Time Comparison}
	\label{fig:time}
	\centering
	\includegraphics[width=0.9\textwidth]{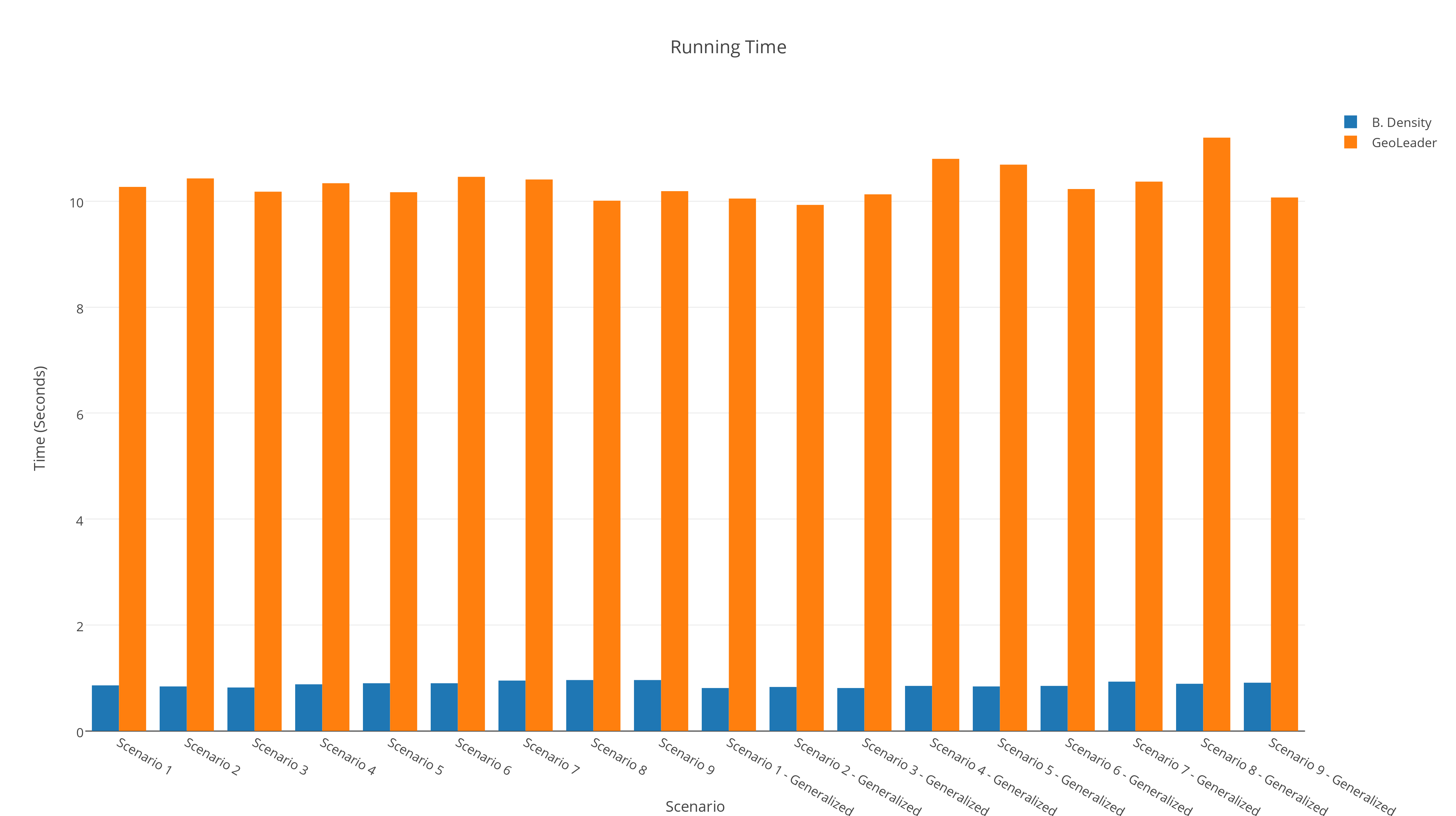}
\end{figure*}

The comparison of the running times shows the largest difference in the measurements taken for the two systems. VBAS has shown significantly faster running times, taking, on average, 8.5\% of the time taken by GeoLeader. A graph of these results can be found in Figure \ref{fig:time}. Due to the need in GeoLeader to evaluate all adjacent regions at each iteration of its process to decide which regions to merge, its process of aggregation ends up being much more lengthy. GeoLeader does not actually guarantee k-anonymity and consequently does not require that equivalence classes be computed. (What it does guarantee is that all aggregated regions have a population above the computed cutoff size.) In contrast, VBAS  spends the majority of its running time on the loading of the records of the data set and the computation of equivalence classes. This is actually the dominant factor in the running time of VBAS. The process of aggregation itself is much faster as there is no need to study the individual regions in order to make decisions about which to merge. The entire aggregation occurs all at once when the Voronoi diagram is constructed and the groups of regions are determined. The loading of records and computing of equivalence classes takes O(nd) time where n is the number of records and d is the number of quasi-identifiers on which we are achieving anonymity. Given d is typically quite low (between 2 and 6), the running time of VBAS essentially scales linearly by the number of records in the data set. As such, VBAS is able to run much more quickly than GeoLeader and is also easily scalable for large data sets.

Another advantage of our system is that it has lower requirements in terms of its input. Both systems require information about an initial regionalization as well as the data set to be anonymized, however GeoLeader requires information about the radii and adjacencies of the initial regionalization which is not necessary for VBAS.

We therefore conclude that VBAS is a preferable choice as a geographic-based system of anonymization as it has lower requirements in terms of input and can produce results of comparable quality with a much faster running time.

\section{Conclusion}
In this paper, we have presented a novel system of geographic-based anonymization for health care data, VBAS. This system runs a process of geographic aggregation guided by the use of a Voronoi diagram. We provide here a single set of approaches that can be applied to the system in order to compare the effectiveness of VBAS to another geographic-based system of anonymization, GeoLeader. Through our tests, we demonstrate that VBAS is able to produce results with comparable quality in terms of the reduction of suppression and information loss. We also show that VBAS provides the advantages of a significantly faster running time and the ability to create slightly more compact regions. Beyond this, VBAS also provides a guarantee of k-anonymity which GeoLeader does not, and has lower requirements for its input. Based on these advantages, we conclude that  VBAS is a preferable choice for the anonymization of data when the preservation of geographic information is a high priority.
		
\section*{Acknowledgment}
The authors gratefully acknowledge financial support from the Natural Sciences and Engineering Research Council of Canada (NSERC) under Grant No. 371977-2009 RGPIN.

\markboth{William Lee Croft, J\"{o}rg-R\"{u}diger Sack and Wei Shi}{International Journal of Geographical Information Science}


\label{lastpage}

\end{document}